\documentclass[preprint,preprintnumbers,amsmath,amssymb,nofootinbib]{revtex4}

\usepackage{mathrsfs}
\usepackage{etex}
\usepackage{amssymb,amsthm,amscd,amsbsy,array}
\usepackage{bm}
\usepackage{amsmath,dsfont}
\usepackage{soul} 
\usepackage{graphics,graphicx,xcolor}

\usepackage{amsthm}

{Corollary}[section]
\newtheorem{prop}
{Proposition}[section]

\usepackage{tikz}
\usetikzlibrary{shapes,calc,arrows,arrows.meta,decorations.markings,angles,math}

\usepackage[colorlinks=true, pdfstartview=FitV, linkcolor=blue, citecolor=blue, urlcolor=blue]{hyperref} 

\newcommand{\blue}{\textcolor{blue}}

\newcommand{\orange}{\textcolor{orange}}
\newcommand{\purple}{\textcolor[rgb]{0.5,0.0,0.5}}
\newcommand{\gb}{\quad\colorbox{green}}

\newenvironment{redtext}{\color{red}}
{\ignorespacesafterend}
\newenvironment{bluetext}{\color{blue}}{\ignorespacesafterend}

\newenvironment{magentatext}{\color{magenta}}{\ignorespacesafterend}
\newenvironment{cyantext}{\color{cyan}}{\ignorespacesafterend}
\newenvironment{orangetext}{\color{orange}}
{\ignorespacesafterend}
\newenvironment{purpletext}{\color{purple}}{\ignorespacesafterend}
\newcommand{\bpurple}{\begin{purpletext}}
\newcommand{\epurple}{\end{purpletext}}

\newcommand{\bmagenta}{\begin{magentatext}}
\newcommand{\emagenta}{\end{magentatext}}
\newcommand{\bcyan}{\begin{cyantext}}
\newcommand{\ecyan}{\end{cyantext}}
\newcommand{\bblue}{\begin{bluetext}}
\newcommand{\eblue}{\end{bluetext}}
\newcommand{\bred}{\begin{redtext}}
\newcommand{\ered}{\end{redtext}}
\newcommand{\borange}{\begin{orangetext}}
\newcommand{\eorange}{\end{orangetext}}

\numberwithin{equation}{section}

\let\ssection=\section
\renewcommand{\section}{\setcounter{equation}{0}\ssection}
\newcommand{\beq}{\begin{equation}}
\newcommand{\eeq}{\end{equation}}
\newcommand{\bec}{\begin{centre}}
\newcommand{\ec}{\end{centre}}

\newcommand{\bb}{{\mathbf{b}}}

\newcommand{\vE}{{\mathbf{E}}}

\newcommand{\bbeta}{\boldsymbol{\beta}}

\newcommand{\bx}{{\bm{x}}}
\newcommand{\bX}{{\bm{X}}}
\newcommand{\bxi}{{\bm{\xi}}}

\def\aand{{\quad\text{\and and}\quad}}

\newcommand{\cE}{{\mathcal{E}}}

\newcommand{\bgamma}{\boldsymbol{\gamma}}

\newcommand{\cK}{{\mathcal{K}}}

\newcommand{\cL}{{\mathscr{L}}}
\newcommand{\cH}{{\mathscr{H}}}

\newcommand{\bp}{{\mathbf{p}}}

\newcommand{\bP}{{\bf P}}

\newcommand{\bQ}{{\bf Q}}

\newcommand{\cT}{\mathcal{T}}

\def\bnabla{{\bm{\nabla}}}

\def\smallover\#1/\#2{\hbox{$\textstyle\frac{\#1}{\#2}$}} %

\def\bp{{\bm{p}}}
\def\bP{{\bm{P}}}

\def\vp{\mathbf{p}}
\def\bequ{\begin{enumerate}}
\def\eenu{\end{enumerate}}
\def\bitem{\begin{itemize}}
\def\eitem{\end{itemize}}
\def\bprop{\begin{prop}}
\def\eprop{\end{prop}}

\def\beq{\begin{equation}}
\def\eeq{\end{equation}}
\def\beqa{\begin{eqnarray}}
\def\eeqa{\end{eqnarray}}
\def\nn{\nonumber}
\def\barray{\left(\begin{array}}
\def\earray{\end{array}\right)}
\def\barraynb{\begin{array}}
\def\earraynb{\end{array}}

\def\?{{\,\gb{\fbox{\texttt{?}}\;}}\,}

\def\p{{\partial}}

\def\vx{\mathbf{x}}
\def\vQ{\mathbf{Q}}

\def\Rarrow{{\quad\Rightarrow\quad}}

\def \p{{\partial}}

\newcommand{\bE}{{\mathbf{E}}}
\def\bP{\mathbb P}

\def\benu{\begin{enumerate}}
\def\eenu{\end{enumerate}}
\def\bitem{\begin{itemize}}
\def\eitem{\end{itemize}}

\def\bP{{\rm{\bf P}}}

\newcommand{\const}{\mathop{\rm const.}\nolimits}
\newcommand{\half }{\smallover{1}/{2}}

\def\smallover#1/#2{\hbox{$\textstyle\frac{#1}{#2}$}} %
\def\smallcirc{{\raise 0.5pt \hbox{$\scriptstyle\circ$}}}
\def\cabove(#1){\stackrel{\smallcirc}{#1}}
\def\ccabove(#1){\,\stackrel{\smallcirc\smallcirc}{#1}\,}
\def\cccabove(#1){\stackrel{\,\smallcirc\smallcirc\smallcirc}{#1}\,}
\def\2{{\smallover1/2}}

\def\boxit#1{
\vbox{\hrule\hbox{\vrule\kern4pt
\vbox{\kern5pt#1\kern5pt}\kern4pt\vrule}\hrule}
} 

\let\ssection=\section
\renewcommand{\section}
{\setcounter{equation}{0}\ssection}

\def\besub{\begin{subequations}}
\def\esub{\end{subequations}}

\begin{document}


\title{
Peierls substitution and Hall motion \\
in exotic Carroll dynamics
\\
}
\author{
H-X. Zeng$^{1}$\footnote{mailto:zenghx53@mail2.sysu.edu.cn},
Q.-L. Zhao$^{1}$\footnote{mailto: zhaoqliang@mail2.sysu.edu.cn},
P.-M. Zhang$^{1}$\footnote{corresponding author.  mailto:zhangpm5@mail.sysu.edu.cn},
and
P.~A. Horvathy$^{2}$\footnote{mailto:horvathy@univ-tours.fr}
}

\affiliation{
${}^{1}$
School of Physics and Astronomy, Sun Yat-sen University, Zhuhai, China
\\
${}^{2}$ Institut Denis-Poisson CNRS/UMR 7013 - Universit\'e de Tours - Universit\'e d'Orl\'eans Parc de Grammont, 37200; Tours, France \\
%
}
\date{\today}

\begin{abstract}
\textit{The particle with first-order dynamics 
  proposed by Dunne, Jackiw and Trugenberger (DJT) to justify the ``Peierls substitution" is
 obtained  by reduction from both of two-parameter centrally extended Galilean and Carroll systems. In the latter case  the extension parameters $\kappa_{exo}$ and $\kappa_{mag}$ generate non-commutativity of the coordinates resp. behave as an internal magnetic field. The position and momentum follow uncoupled anomalous Hall motions. Consistently with  partial immobility, one of the Carroll boost generators is broken but the other remains a symmetry. Switching off $\kappa_{exo}$, the immobility of unextended Carroll particles is recovered. The Carroll system is dual to an uncharged anyon on the horizon of a black hole which exhibits the spin-Hall effect.}
 \bigskip
 
\noindent
 Phys. Rev. D \textbf{111} (2025) no.4, 044025
doi:10.1103/PhysRevD.111.044025
[arXiv:2411.14329 [hep-th]].

\end{abstract}

\maketitle

\tableofcontents


\section{Introduction
}\label{Intro}

The fundamental property which has long delayed interest in Carrollian physics is that \emph{particles with Carroll symmetry can not move} \cite{Leblond,SenGupta}.
In this paper we show that
 a particle associated with the  recently discovered
  ``exotic'' \emph{two-parameter central extension of the planar Carroll group} \cite{Azcarraga,Ancille0,Ancille1,Ancille2,Marsot21} analogous to the one of the planar Galilei group in \cite{LLGal}
\emph{does  move}, namely by following the Hall law. However we confirm that turning off the exotic extension a massive Carroll particle does indeed not move.  

Carroll symmetry  is  highlighted by the curious action of Carroll (or C-) boosts which leave the position fixed and act only on ``Carroll time'' we denote here by $s$ \cite{Leblond,SenGupta}, 
\beq
\bx \to \bx,
\qquad
s \to s - \bb\cdot\bx\,,
\label{Cboost}
\eeq
to be compared with usual Galilei boosts,
\beq
\bx \to \bx + \bb\,t\,, \qquad t \to t\,.
\label{Gboostbis}
\eeq
The  sign difference between Galilei and Carroll boosts is conveniently explained  in the Kaluza-Klein-type ``Bargmann'' framework \cite{DBKP,DGH91,Carrollvs}. The  extended space time has coordinates $\bx,t,s$ where $t$ and $s$ are lightlike. Galilean physics is a projection onto non-relativistic spacetime with coordinates $(\bx,t)$, with $s$ corresponding to its central extension \cite{Bargmann}. The arena of Carrollian physics is in turn the restriction of extended space to $t=\const$ and has thus coordinates $(\bx,s)$.

Carrollian physics has become lately a much discussed topic \cite{Gibbons02,Carrollvs,Bergshoeff14,Morand1,Morand2,Marsot21,deBoer,deBoer23}, justified by its applications to gravitational waves, dark matter and to black hole physics, see  \cite{Carroll4GW,Laura19,Laura22,Laura23,MZH,Gray,Aggarwal,Ecker23,Ecker24,MZCH} and many further references. Our new results in this paper complete previous work  \cite{Leblond,deBoer,Casalbuoni23,MultiCarroll}.

We start with recalling  the \emph{Peierls substitution} proposed many years ago in many-body physics \cite{Peierls, Luttinger,Kohn,Azbel,Hofstadter}~:
for a charged particle in the plane subjected to  perpendicular magnetic and electric fields  and constrained to stay in the lowest Landau level (i) the components of quantum position operator should be \emph{non-commuting} and (ii) its Hamiltonian should reduce to the mere potential and have \emph{no kinetic term},
\beq
\big[{\widehat{\bxi}}_m, \widehat{\bxi}_n\big]=\frac{i}{eB}\epsilon_{mn} \aand \widehat{\cH} = eV(\widehat{\bxi})\,. 
\label{noncommrel}
\eeq
Half of a century later, Peierls' idea was resurrected by Dunne, Jackiw and Trugenberger (DJT) \cite{DJT,DunneJP}. 
A remarkable feature of their first-order equations 
is that the motion obeys the \emph{Hall law}, \eqref{Halllaw} below. 

In \cite{DHPeierls} we re-derived the
DJT system from the non-commutative mechanics associated with the exotic (double) extension of the planar Galilei group noticed by  L\'evy-Leblond \cite{LLGal} and further investigated in \cite{Kosinski,Grigore,LSZ,DHPeierls,DHJPA,NCLandau,ExoRev,ZHChiral,PlyushChiral1,PlyushChiral2} by Hamiltonian reduction \cite{FaTMF,FaJa}. Our new paper carries out a similar study in
the hitherto unexplored exo-Carrollian case, recalling occasionally the Galilean one for the sake of comparison.
\bigskip

We proceed in three steps. After recalling the  DJT system (sec.\ref{DJTSec}) we 
show that it can in fact be derived  from both exotic Galilei \emph{and}  exotic  Carroll particles. Then focusing our attention at the latter, we show that a charged exotic Carroll  particle \emph{moves} by following an \emph{anomalous Hall law}  \eqref{aCHall} \cite{Ezawa,Stone} both in the presence, but also in the \emph{absence} of an external magnetic field, reminiscent of the anomalous Hall effect \cite{Kar+,AnomHall}. 
Turning off the  exotic extension we recover an usual Carroll particle with its celebrated immobility. 

For simplicity, we limit our attention to constant electromagnetic fields $B,\, \bE$ in the plane. Since we work in flat euclidean space, vectors will have lower indices.

\section{The Dunne-Jackiw-Trugenberger (DJT) system
}\label{DJTSec}

In the early nineties Dunne, Jackiw, and Trugenberger (DJT)  \cite{DunneJP,DJT} reconsidered the Peierls Ansatz \cite{Peierls},  
 arguing that for a charged particle in the plane, letting the mass go to zero {turns off the kinetic term}, leaving us with the \emph{first-order-in-time-derivative} Lagrangian which will hence be called that of a DJT  particle,
 \beq
\cL_{DJT} =
\frac{eB}{2}\,
\epsilon_{ij}{\xi}_i{\xi}_j^{\prime}
\; - \;eV(\bxi)\,, 
\label{DJTlag}
\eeq
where the prime means derivation w.r.t. Galilean time $t$, $\{\cdot\}'=d/dt$. $B=\const\neq0$ and $\bE=\const$ will hence be assumed for simplicity.
The electric charge drops out from the equations of motion as long as it does not vanish, providing us with~:
\begin{prop}
{The DJT particle moves by following the Hall law} \cite{DHPeierls,DHJPA},
\beq 
{\xi}_i^{\prime} = \epsilon_{ij}\frac{E_j}{B\;}\,.
\label{Halllaw}
\eeq\vskip-3mm
\label{DJTHallP}
\end{prop}
Legendre transformation of  
 \eqref{DJTlag} then yields the Poisson brackets and Hamiltonian,
\beq
\Big\{{\xi}_i,{\xi}_j\Big\}= -\frac{1}{eB\,}\epsilon_{ij}
\aand \cH=eV(\bxi)\,,
\label{DJTHamstr}
\eeq
respectively. 
Here $B$ is viewed as a fixed external quantity.
 The Poisson bracket corresponds to the symplectic form \cite{DHPeierls,DHJPA},
\beq
\Omega_{DJT}=
\half{eB}\epsilon_{ij}d{\xi}_i \wedge d{\xi}_j\,.
\label{DJTsymp}
\eeq

To sum up,
\begin{itemize}
\item 
The DJT Lagrangian \eqref{DJTlag}  has no mass term  and is of the first order in the velocity, implying first-order equations of motion, \eqref{Halllaw};
\item
The coordinates do not commute, \eqref{DJTHamstr}, 
and the {Hamiltonian} is the \emph{potential} with {no kinetic term}; 

\item  
The Hall law \eqref{Halllaw}
 implies \emph{partial immobility}:
 the particle moves perpendicularly to the electric field,
 $\bE = -\bnabla V$. When the electric field  vanishes,
 $\bE = 0$, then the DJT particle \emph{does not move} at all,
 $ \bxi = \bxi_0=\const$ 

to be compared with the usual circular motions of a charged massive particle in a magnetic field. %

\end{itemize}
 \goodbreak

We note for later use that that
the motions can also be obtained by finding the kernel of the Souriau 2-form \cite{SSD,Marsot21},
\beq
\sigma_{DJT} =  d\cL_{DJT}\wedge dt= 
\half{eB}\epsilon_{ij}d{\xi}_i \wedge d{\xi}_j + eE_i d{\xi}_i \wedge dt\,,
\label{DJTsigma}
\eeq
see the Appendix \ref{Appendix}.

In the next section  we show tht the  DJT system  can be obtained  by reduction from \emph{two} different, centrally extended  particle models.
After recalling some relevant facts from the Galilean case for comparison, we carry out a similar but more detailed analysis for Carroll.


\section{Exotic Galilean dynamics}\label{exoGal}

As shown in \cite{DHPeierls,DHJPA,NCLandau,ExoRev,ZHChiral},
 the DJT system is obtained  from ``exotic'' Galilean mechanics \cite{Kosinski,Grigore,LSZ,PlyushChiral1,PlyushChiral2}  associated with  the \emph{two-parameter central extension of the planar Galilei group} \cite{LLGal} by Hamiltonian reduction \cite{FaTMF,FaJa}. 
One of the parameters present in any dimension is  the mass, $m$ \cite{Bargmann}. The other, ``exotic'' one we denote here by $\kappa_{exo}$ is specific for planar physics \cite{LLGal}. 
 In terms of the non-commutativity parameter and the effective mass,
\beq
\theta=\frac{\kappa_{exo}}{m^{2}} \aand m^*=m(1-e{\theta}{B})\,,
\label{Geffmass}
\eeq
the motion in a planar electromagnetic field  $\vE,B$
[assumed constant for simplicity]  is described by the equations
\cite{DHPeierls,DHJPA,NCLandau,ZHChiral,ExoRev,Zhang:2011kr,PolPer}, 
\beqa
m^*x_i^{\prime}=p_i-me\theta\epsilon_{ij}E_j,
\qquad
p_i^{\prime}=eB\epsilon_{ij}{x}_j^{\prime}+eE_i\,.
\label{Gexoeqmot}
\eeqa

When $m^*\neq0$ the system is regular; the eqns. (\ref{Gexoeqmot}) derive from the Hamiltonian and symplectic form, 
\besub
\begin{align}
\cH_{exoG}&=\frac{\vp^{2}}{2m}+eV(\vx)\,,
\label{GHamil}
\\[3pt]
\Omega_{exo} &=
dp_i\wedge dx_i
+\frac{\theta}{2}\epsilon_{ij}dp_i\wedge dp_j
+\frac{eB}{2}\epsilon_{ij}dx_i\wedge dx_j\,,
\label{exoSymp}
\end{align}
\label{GSH}
\esub
by bracketing with  the exotic Poisson brackets,
\beqa
\{x_i,x_j\}=\frac{\theta}{1-e{\theta}{B}}\,\epsilon_{ij},
\quad
\{x_i,p_j\}=\frac{1}{1-e{\theta}{B}}\,\delta_{ij},
\quad
\{p_i,p_j\}=\frac{eB}{1-e{\theta}{B}}\,\epsilon_{ij}\,
\label{GexoPB}
\eeqa
or alternatively, 
from the kernel of the exotic Souriau form  $\sigma=\Omega-d\cH\wedge dt$ cf.  \eqref{sigmaOmegaH}, 
\beq
\sigma_{exoG}= dp_i\wedge dx_i +\frac{\theta}{2}\epsilon_{ij}dp_i\wedge dp_j
+\frac{eB}{2}\epsilon_{ij}dx_i\wedge dx_j
+eE_{i}dx_{i}\wedge dt-\frac{p_{i}}{m}dp_{i}\wedge dt\,.
\label{eGalsigma}
\eeq 

Eqns \eqref{Gexoeqmot} can be solved analytically 
\cite{DHJPA,ZHChiral}~:

\bprop
When the effective mass  does not vanish, $m^*\neq0$, the motions combine  rotation and Hall drift,
\begin{eqnarray}
    	&&
    	x_1(t)=\frac{c_4}{eB}\cos\left(\frac{eB}{m^*}\,t\right) + \frac{c_3}{eB}\sin\left(\frac{eB}{m^*}\,t\right) + \frac{E_2}{B}t + c_2\,, 
		\\[4pt]
    	&&
    	x_2(t)=\frac{c_3}{eB}\cos\left(\frac{eB}{m^*}\,t\right) - \frac{c_4}{eB}\sin\left(\frac{eB}{m^*}\,t\right) - \frac{E_1}{B}t + \frac{m\,E_2}{e\,B^2} + c_1\,,
    	\\[4pt]
    	&&
    	p_1(t)=\cos\left(\frac{eB}{m^*}t\right)\,c_3 - \sin\left(\frac{eB}{m^*}t\right)\,c_4 + \frac{m\,E_2}{B}\,,
    	\\[4pt]
    	&&
    	p_2(t)=-\cos\left(\frac{eB}{m^*}t\right)\,c_4 - \sin\left(\frac{eB}{m^*}t\right)\,c_3 - \frac{m\,E_1}{B}\,.
    \end{eqnarray}
The arbitrary constants $c_i$  fix the initial position, and   
the radii of the circles.
\label{regGalP}
\eprop

It was shown in earlier work \cite{DHPeierls,DHJPA,NCLandau,ExoRev} that
when the magnetic field takes the critical value 
\beq
B=B_{crit}=\frac{1}{e\theta}\,,
\label{GcritB}
\eeq
or equivalently when the effective mass vanishes, $m^*=0$,  we have ~: 
 
\bprop
In the critical case \eqref{GcritB} the momentum is  determined  by the field. Both
 the position coordinate $\bx$ 
 moves according to the Hall law. 
\beq
p_i=m\,\theta\epsilon_{ij}eE_j=\const\,
\aand
{x}_i^{\prime}=
\epsilon_{ij}\frac{E_j}{B_{crit}}\,. 
\label{GHalllaw}
\eeq
\label{Gcrit1P}
\eprop \vskip-6mm
More insight is obtained by combining the position and the momentum into the \emph{guiding centre} \cite{Ezawa,Stone,ZHChiral},
\begin{equation}
Q_{i}=x_{i}+\displaystyle\frac{1}{eB}\epsilon_{ij}p_{j}\,,  
\label{Bguiding}
\end{equation}
for which Proposition \ref{regGalP} allows us to deduce~:
\begin{prop}
When $m^*\neq0$, then the guiding centre moves, \underline{independently} of the \underline{non-commutativity parameter $\theta$} and even outside the critical value \eqref{GcritB}, according to the Hall law,
\beq
Q_{i}=\epsilon_{ij}\frac{E_j}{B}\,t 
+ C_i\, \qquad C_i=\const\,
\label{GQHall}
\eeq
\label{Gguiding}
\eprop\vskip-6mm 

The trajectories shown in FIG. \ref{Exotic-G} are circles which drift along the guiding centre which itself follows the Hall law \eqref{GQHall} consistently with   FIG.s \#3 and \#4 of \cite{ZHChiral}. 
\begin{figure}[h]
\includegraphics[scale=.311]{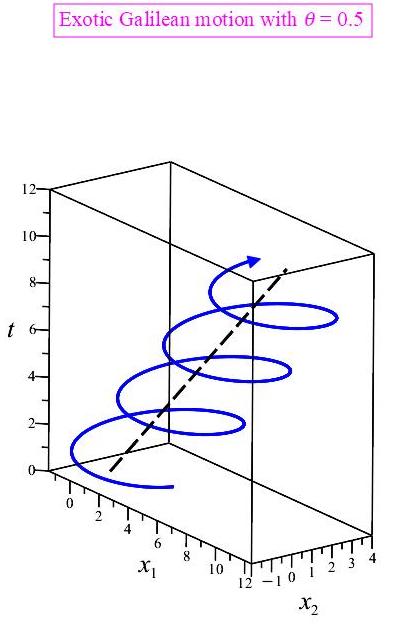}\quad
\includegraphics[scale=.3]{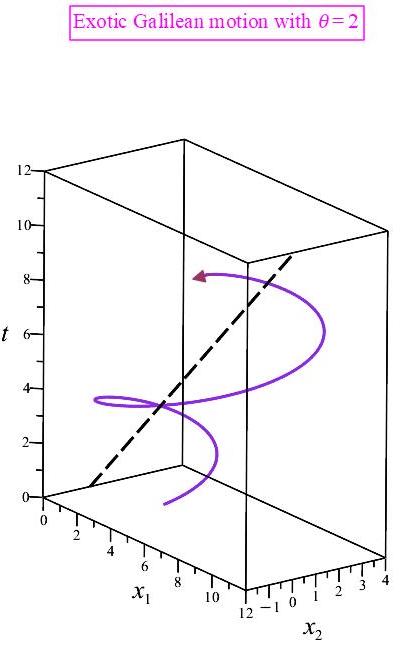}\quad
\includegraphics[scale=.32]{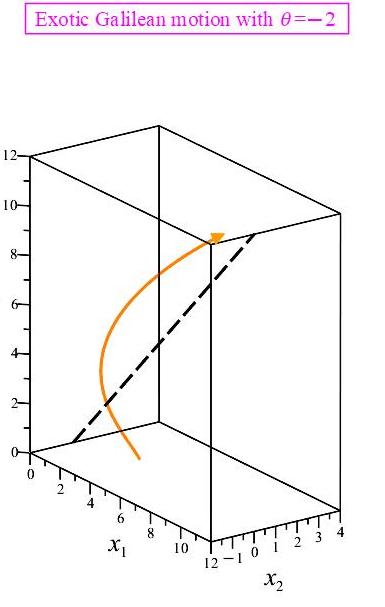}\\[6pt]
\includegraphics[scale=.35]{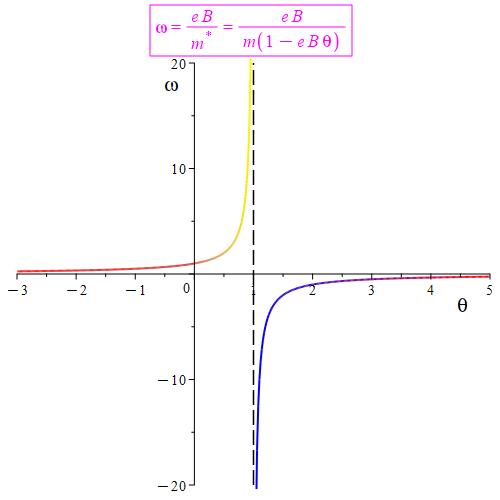}
\vskip-2mm
\caption{\textit{\small
(i) Exotic Galilean motions with parameters $e=1,\, B = 1,\, \bm{E}=(0,1),\, m = 1,\, c_1=c_2=c_3=0,\, c_4=3$. The \blue{\bf blue}/\purple{\bf purple} curves are below (\blue{$\theta=0.5$}) / above (\purple{$\theta=2$}) the critical value $\theta=1$.  
The \orange{\bf orange} curve is for large $|\theta|$ (\orange{$\theta=-2$)}. (ii)  When $\theta\to 1$ either from the left or from the right, the frequency diverges and changes sign.  
}
\label{Exotic-G}
}
\end{figure}

When the critical case  \eqref{GcritB} 
 is approached, the rotation speeds up and  its \emph{angular velocity}  
$\omega^*={eB}/{m^*}$ diverges. Crossing the critical value the direction of the rotation is reversed.

For $|\theta|$ very large instead the effective mass diverges, $m^* \to \infty$, the frequency tends to $0$ and the motion reduces to the (displaced) guiding centre,
\beq
x_1(t)=
 \displaystyle\frac{E_2}{B}t + \const 
\qquad
x_2(t)=
-  \displaystyle\frac{E_1}{B}t + \const
\label{bigtheta}
\eeq
shown by the orange curve in FIG.\ref{Exotic-G}.

Further progress is achieved by
rewriting the Souriau form \eqref{eGalsigma} in terms of $Q_i$,   
\beqa
&\sigma_{exoG}= \sigma_{QG} + \sigma_{pG} = \qquad 
\label{GsigmaQ}
\\[8pt]
&\Big\{\half eB \epsilon_{ij}dQ_i\wedge dQ_j +\, eE_i dQ_i\wedge dt\Big\}
+ 
\Big\{\half\big(\theta - \frac{1}{eB}\big) \epsilon_{ij}dp_i\wedge dp_j 
+ 
\big(\epsilon_{ij}\frac{E_j}{B}-\displaystyle\frac{p_i}{m}\big) dp_i\wedge dt\Big\}
\nn
\eeqa
where the first  describes the (Hall) dynamics of the guiding centre \eqref{GQHall}, and $\sigma_{pG}$ determines that of the momentum which however intervenes in real-space dynamics. 

When the magnetic field takes the critical value in \eqref{GcritB}
or equivalently, when the effective mass vanishes, $m^*=0$, then  
Hamiltonian reduction 
shows  \cite{DHJPA,NCLandau} that the phase 4-dimensional space reduces to a 2-dimensional one
 upon which the guiding centre  \eqref{Bguiding} is a good coordinate. 
The reduced Hamiltonian structure
\beq
\Omega_{red}= \frac{1}{2\theta}d\bQ\times d\bQ 
\quad\;\Leftrightarrow\quad\;
\Big\{Q_{1},Q_{2}\Big\} = -\theta =-\frac{1}{eB_{crit}}
\,,
\quad  
\cH_{red}=eV(\vQ)\,
\label{redHamstr}
\eeq
or with the ``Chern-Simons-type" Lagrangian of \cite{DunneJP,DJT}
with no kinetic term,
\begin{equation}
\cL_{red}=\frac{1}{2\theta}\vQ\times{\vQ}^{\prime}-V(\vQ)\,,
\label{redlag}
\end{equation}
Comparing with the DJT expressions \eqref{DJTsymp} and  \eqref{DJTlag},
 we recover the DJT system in section \ref{DJTSec} with 
the guiding centre $\bQ$ replacing $\bxi$. 

For later comparison, the reduced Souriau form is
\beq
\sigma_{exoG}^{red}=\half eB \epsilon_{ij}dQ_i\wedge dQ_j +\, eE_i dQ_i\wedge dt.
\label{xoGr}
\eeq
  
 \section{Exotic Carroll particle}\label{exoCarr}
 
 A remarkable fact which has long escaped attention and was recognised only recently \cite{Azcarraga,Ancille0,Ancille1,Ancille2,Marsot21} is that, unlike in $d\geq3$ dimensions, the \emph{planar Carroll group} admits also a \emph{two-parameter central extension} 
The  associated particle will be called an \emph{exotic Carroll particle}. 
(Our notations correspond to $\kappa_{exo} = m^{2}\theta =2q_2,\;
 \kappa_{mag}=-2q_1$ in \cite{Marsot21}.)
We assume that our particle has non-vanishing mass and  electric charge, $m\neq0$ and $e\neq0$, respectively, and is subjected to constant  planar magnetic and electric fields. Unlike as for Galilei \cite{Bargmann},  the mass $m$ is for Carroll an externally given constant and  {not} a central-extension parameter.   

In terms of  the {non-commutativity parameter}  and the {effective magnetic field}, 
\beq
\theta = \frac{\kappa_{exo}}{m^{2}}
\aand
B^* = eB + \kappa_{mag}\,, 
\label{B*field}
\eeq
respectively, Souriau's orbit construction \cite{SSD}
applied to the exotic Carroll group yields the Souriau form
 \cite{Ancille0,Ancille1,Ancille2,Marsot21},
\beq
\sigma_{exoC} = \underbrace{d\bp\wedge d\bx
+ 
\frac{{B^*}}{2}\,\epsilon_{ij}dx_i\wedge dx_j
+
\frac{\theta}{2}\,\epsilon_{ij}dp_i\wedge dp_j
}_{\Omega_{exo}} + e E_i\, dx_i\wedge ds\,.
\label{eCarrsigma}
\eeq

Comparison with \eqref{eGalsigma} shows that (up to $t \leadsto s$ and $eB \leadsto B^*$) this is identical to
 the Galilean expression \emph{except for the missing kinetic term in  
 the Hamiltonian} which is merely the ``naked'' potential,
\beq
{\cH}_{Car} = eV\,.
\label{CHamilt}
\eeq
The dynamics is described alternatively by the doubly-extended first-order-in-$\dot{\bx}$ phase-space Lagrangian
 with no kinetic term \cite{MZH,MZCH},
\beq
\cL_{Car} = \bp\cdot\dot{\bx}
+
\half B^*\epsilon_{ij}x_i\dot{x}_j
+
\half \theta 
 \,\epsilon_{ij}p_i\dot{p}_j
- eV\,,
\label{exoCLag}
\eeq
where the ``dot'' means derivation w.r.t. Carroll time, $d/ds$. 
The absence of the kinetic term implies that the Carroll equations of motion are {decoupled} for all values of the parameters \cite{Marsot21}, 
\beq
\big(1-\theta B^*\big)
\dot{x}_i = - e\theta\epsilon_{ij}E_j
\;\aand\;
\big(1-\theta B^*\big)\dot{p}_i = eE_i\,
\label{Cmot}
\eeq 
Then introducing again  the effective mass, 
\beq
m^* = 1-\theta B^*\,
\eeq
cf. \eqref{Geffmass} we conclude~:
\begin{prop}
 When $m^* \neq 0$ the system is regular  and
 the coordinates follow an {anomalous Hall law} \cite{Ezawa,Stone} 
 \beq
\dot{x}_i = -\big(\frac{e\theta}{1-\theta B^*}\big)\,
 \epsilon_{ij} E_j  \;\Rarrow\;
 x_i(s) = - \big(\frac{e\theta}{1-\theta B^*}\big)\,\epsilon_{ij}{E_j}s  + x_i^0\,, 
\label{aCHall}
\eeq
 supplemented  by
\beq
\dot{p}_i = \big(\frac{1}{1-\theta B^*}\big)eE_i
\;\Rarrow\;\;
p_i(s) = \big(\frac{eE_i}{1-\theta B^*}\big)\, s + p^0(s)\,.\qquad\;
\label{pmotion}
\eeq
Switching off  the exotic parameter, $\theta=0$, the immobility of Carroll particles is duly recovered.
\label{anomCHallP}
\end{prop}
The $p$-equation has no effect on the dynamics of the position, though, and could actually be ignored when studying the $\bx$-dynamics.
We recall for comparison that Galilean particles follow the Hall law  \eqref{GHalllaw}  {only in the  critical case} \eqref{GcritB} when the effective mass vanishes.
\goodbreak

The anomalous Hall trajectories for various values of $m^*$ are depicted  in FIG.\ref{xv-theta}(i). 

\begin{figure}[h]
\includegraphics[scale=.45]{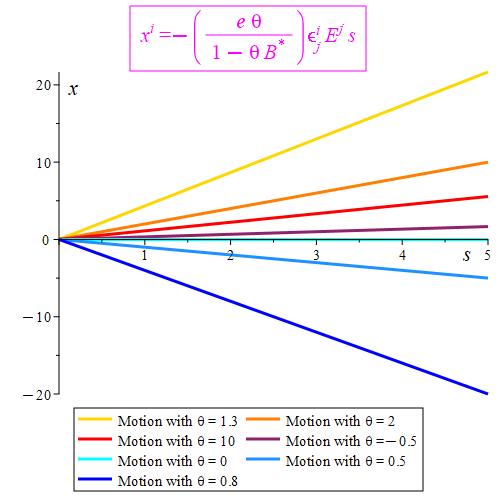}\quad
\includegraphics[scale=.45]{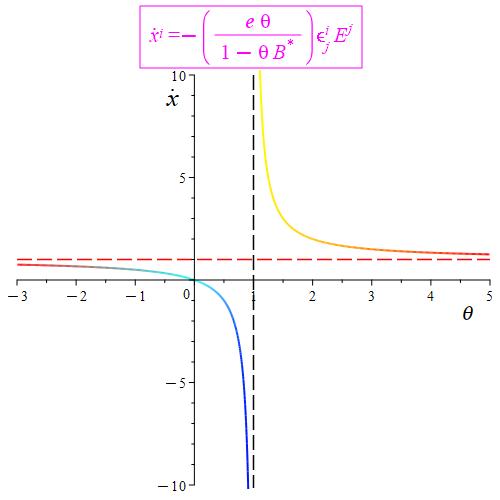}
\\
(i) \hskip 70mm (ii)
\vskip-3mm
\caption{\textit{\small (i) Hall motions 
 for various values of the non-commutativity parameter $\theta\neq 1$, after fixing the other parameters as $e=1,\, B^* = 1,\, \bE=(0,1)$ so  that the motion remains  perpendicular to the electric field $\bE$.  
For $\theta=0$ we recover the unextended Carroll situation with no motion.  (ii)  When $\theta\to 1$ the velocity diverges. For very large $\theta$ instead, the residual velocity tends to the Hall value $\dot{x}_{1}=eE/B^*$ (to be compared with the Galilean case in FIG.\ref{Exotic-G}).
}
\label{xv-theta}
}
\end{figure}
\goodbreak
Approaching the critical value
\beq
m^*=0 \quad \Leftrightarrow\quad  \left({\theta}{B^*}\right)_{crit} = {1}\,,
\label{Carrcrit}
\eeq 
the velocity diverges, cf. FIG.\ref{xv-theta}(ii).
The coefficient in \eqref{aCHall} changes sign when the denominator crosses zero, flipping over the direction of the motion, as  in FIG.\ref{xv-theta}ii. 
 --- to be compared with the reversal  of the rotation in exotic Galilean mechanics \cite{ZHChiral} in FIG.\ref{Exotic-G}. The novelty is that we get Hall-like Carroll motion \emph{always}, not only in the critical case. 

An interesting particular case is obtained by presenting  \eqref{aCHall} as,
\beq
\big(\underbrace{1 - \kappa_{mag}\theta}_{0} - e \theta{B}\big)\dot{x}_i = 
- e\theta\epsilon_{ij} E_j\,.
\label{aCHallbis}
\eeq
Then putting  the underbraced quantity  to zero,  
\beq
\theta\kappa_{mag}=1
\quad\Leftrightarrow\quad
\kappa_{exo}\kappa_{mag}=m^{2}\,.
\label{thetakappa1}
\eeq
Then the exotic factor $e\theta$ drops out (as long as it does not vanish), yielding~:

\begin{prop}
When the parameters satisfy \eqref{thetakappa1} eqn.
 \eqref{aCHall} reduces to the usual Hall law,
\beq
\dot{x}_i = \epsilon_{ij} \frac{E_j}{B} \,.
\label{CarrHall}
\eeq
\label{uCHallP}
\end{prop}\vskip-6mm
Eqn. \eqref{aCHallbis} suggests to consider yet another remarkable particular case:

\begin{prop}
When the external magnetic field  is switched off, 
\beq
B=0\,,
\label{B0}
\eeq
then, assuming that  
$ \, \theta\,\kappa_{mag} \neq 1,$  the position satisfies
\beq
\dot{x}_i = -\frac{e\theta}{1-\theta\kappa_{mag}}
\epsilon_{ij}E_j\,.
\label{anomHallef}
\eeq
\label{anomHallP}
\end{prop} \vskip-6mm\noindent
Thus when $\theta\neq0$ Hall motions persist even for
$\kappa_{mag}=0$.  \eqref{anomHallef} is
 reminiscent of ``the'' \emph{anomalous Hall effect} observed in ferromagnetic materials in the \emph{absence of an external magnetic field}  \cite{Kar+,Ezawa,Stone,AnomHall}.  
Comparison of \eqref{anomHallef} with \eqref{aCHall} underlines that $\kappa_{mag}$ behaves indeed as a proper magnetic field carried by the particle, and could  be viewed as the planar analog of
 a magnetic monopole. It is also reminiscent of the rather mysterious magnetic field of relativistic anyons \cite{FeherAny}.
 
We underline that the system is regular in both of these particular cases
\eqref{thetakappa1} and \eqref{B0}.

\subsection{The singular Carroll case}\label{CsingSec}

In the singular case \eqref{Carrcrit} both equations in \eqref{Cmot} are
 contradictory unless $e\bE=0$, when they are undetermined.
 This is unlike as in exotic Galilean mechanics,
eqn. \eqref{Gexoeqmot}, where the $p$-terms on the rhs can compensate the exotic one, allowing, for  vanishing effective mass, 
motions which follow the Hall law, \eqref{GHalllaw}
 \cite{DHPeierls,DHJPA,NCLandau,ExoRev,ZHChiral}.
 
We resort to Hamiltonian reduction \cite{FaTMF,FaJa,DHJPA,NCLandau}   
along the lines as for exotic Galilei \cite{DHJPA,NCLandau}.  
We start with the regular case $m^*\neq0$ and introduce new, canonical coordinates \cite{DHJPA},  
\besub
\begin{align}
	Q_i&=x_i+\frac{1}{B^*}\left[1-\sqrt{\frac{m^*}{m}}\right]\,\epsilon_{ij}p_j\,,
\label{Qxp}	
	\\[4pt]
	P_i&=\sqrt{\frac{m^*}{m}}\,p_i-\half B^* \epsilon_{ij} Q_j\,,
	\label{PpQ}
\end{align}
\label{QP}
\esub
in terms of which the symplectic form  and Hamiltonian are, 
\besub
\begin{align}
\Omega_{exo} &=dP_i\wedge dQ_i \,, 
\\[6pt]
\cH_{exoC} &=-\frac{1}{2}\left(\sqrt{\frac{m}{m^{*}}}+1\right)
eE_{i}dQ_{i} 
+\frac{1}{B^{*}}\left(\sqrt{\frac{m}{m^{*}}}-1\right) \epsilon_{ij}eE_{i}dP_{j}\;.
\end{align}
\label{PQOH}
\esub
The corresponding Poisson brackets and equations of motion are
\begin{equation}
\left\{Q_{i},P_{j}\right\} =\delta_{ij},\text{ \ }\left\{
Q_{i},Q_{j}\right\} =0,\text{ \ }\left\{P_{i},P_{j}\right\} =0
\end{equation}%
and 
\begin{equation}
\dot{P}_{i}=\frac{1}{2}\left(\sqrt{\frac{m}{%
m^{*}}}+1\right) eE_{i}
\aand
 \dot{Q}_{i}=-\frac{1}{B^*}\left(\sqrt{\frac{m%
}{m^{*}}}-1\right) \epsilon_{ij}\,eE_{j}\,.
\end{equation}
respectively.
In the critical case \eqref{Carrcrit} the inverse transformation,
\beq
x_{i} =\frac{1}{2}\left(\sqrt{\frac{m}{m^{*}}}+1\right) Q_{i}-\frac{1}{B^{*}}%
\left(\sqrt{\frac{m}{m^{*}}}-1\right) \epsilon_{ij}P_{j}\,,
\eeq
becomes singular \eqref{Carrcrit}, the velocities diverge, however  
 the coordinates \eqref{QP} do have a well-defined limit. Letting  $m^*\to 0$,  \eqref{PpQ} implies that $\bP$ is rotated rigidly by $\pi/2$ w.r.t.  
$\bQ$,
\beq
P_i=-\half B_{crit}^*\epsilon_{ij}Q_j\,,
\label{PfixQ}
\eeq 
leaving us with the Carroll guiding centre,
\beq
Q_i=x_i+\frac{1}{B^*}\,\epsilon_{ij}p_j\,,
\label{Cguiding}
\eeq
as only dynamical degree of freedom.   
The reduced symplectic form, Poisson brackets and reduced Hamiltonian are,%
\begin{equation}
\Omega =\frac{1}{2}B^*_{crit}\epsilon_{ij}dQ_{i}\wedge dQ_{j}\,,
\qquad
\big\{Q_{1},Q_{2}\big\} =-\frac{1}{B^*_{crit}}\,,
\qquad 
\cH_{red}=-eE_{i}Q_{i}\,
\label{redOPB}
\end{equation}
respectively,  cf. \eqref{Bguiding}.
It follows that~: 
\begin{prop} 
In the critical case \eqref{Carrcrit}  the guiding centre $Q_i$ follows  the Hall law \eqref{redHamstr}, 
\beq
\dot{Q}_i = \epsilon_{ij}\frac{E_j}{B^{*}_{crit}}\,,
\label{CQHall}
\eeq
whereas $P_i$ is determined algebraically by $Q_i$ as in \eqref{PfixQ}.
\label{GDCHallP}
\end{prop}

\subsection{Chiral decomposition of exotic Carroll dynamics}\label{chiralSec}

Further  insight is provided by chiral decomposition  proposed in  \cite{PlyushChiral1,PlyushChiral2}. 
Starting with the exotic Souriau form $\sigma_{exoC}$ \eqref{eCarrsigma} (or symplectic form $\Omega_{exo}$ \eqref{exoSymp} and Hamiltonian, $\cH_{Car}$  \eqref{CHamilt}),
we introduce chiral coordinates, 
\begin{equation}
p_{i}=B^*\epsilon_{ij}X^{-}_{j}\,, \qquad
x_{i}=X^{+}_{i}+X^{-}_{i}\,.
\label{X+X-}
\end{equation}%
Here $X^{+}_{i}=Q_i$ 
is in fact the guiding centre, \eqref{Cguiding}. 
The symplectic form and Hamiltonian are decomposed as, %
\besub
\begin{align}
\Omega =\, & \Omega^{+}+\Omega^{-} \;\; \;=
\Big\{\frac{B^*}{2}\epsilon_{ij}dX^{+}_{i}\wedge dX^{+}_{j}\Big\}
-
\Big\{\frac{B^*}{2}(1-{\theta}B^*)\epsilon_{ij}dX^{-}_{i}\wedge dX^{-}_{j}\Big\},
\label{chirO}
\\[6pt]
\cH = \,& \cH^{+}+\cH^{-}=-eE_{i}X^{+}_{i}-eE_{i}X^{-}_{i}\,.
\label{chirH}
\end{align}
\label{chirdec}
\esub%
We note the relative minus in \eqref{chirO} but not in \eqref{chirH}.

In the regular case $m^*= (1-B^*\theta)\neq0$
the associated Poisson brackets%
\begin{equation}
\left\{X^{+}_{i},X^{+}_{j}\right\} =-\frac{1}{B^*}\epsilon_{ij},\text{ \ \ 
}\left\{ X^{-}_{i},X^{+}_{j}\right\} =0,\text{ \ \ }\left\{
X^{-}_{i},X^{-}_{j}\right\} =\frac{1}{B^*\left(1-{\theta}B^*\right)}%
\epsilon_{ij}\,,
\label{exoGchiralPB}
\end{equation}%
yield the  Hamilton equations,%
\besub
\begin{align}
&\frac{dX^{+}_{i}}{ds}= \quad\; %
\epsilon_{ij}\frac{e E_{j}}{B^*}\,,\quad
\label{X+motion}
\\[6pt]
&\frac{dX^{-}_{i}}{ds}  
= \;-\frac{1}{\left(
1-\theta{B^*}\right)}\epsilon_{ij}\frac{e E_{j}}{B^*}\,.
\label{X-motion}
\end{align}
\label{Cchireq}
\esub
which imply~:
\bprop
Both chiral coordinates follow a Hall law --- the guiding centre, $X^{+}_{i}$,  the usual one and  $X^{-}_{i}$ an anomalous one which involves also the exotic parameter $\theta$. 
\label{CchiralmotionP}
\eprop

For $m^* > 0$  $X^{-}_{i}$  moves in the opposite direction as $X^{+}_{i}$ does but  switches  direction beyond the critical value.  
For comparison, remember that in the Galilean case  $X^{-}_{i}$ circles around the guiding centre $X^{+}_{i}$ \cite{ZHChiral}. Combining $X^{+}_{i}$ and $X^{-}_{i}$ the anomalous Hall dynamics \eqref{aCHall} is recovered. 
Eqn. \eqref{Cchireq} allows us to deduce~: 

\begin{prop}
When the non-commutative parameter is switched off then $X^{-}_{i}$  cancels the  motion of the guiding centre $X^{+}_{i}$ and
the particle's position coordinate, $\bx$, stops moving,
\begin{equation}
\theta =0
\Rarrow
\dot{\bx} =\dot{X}^{+} + \dot{X}^{-} = 0\,. 
\label{noxmotion}
\eeq 
 In other regular cases with  $\theta\neq0 $ the motion is a 
combination of  Hall motion of $X^{+}_{i}$ deviated by that of $X^{-}_{i}$,
as illustrated in FIG.\ref{xv-theta}.
\label{nomotP}
\end{prop}
 \goodbreak

The same conclusion could have  been derived directly by combining  $\bx$ and $\bp$ into 
\beq
{\widetilde{Q}}_i = x_i + \theta \epsilon_{ij}p_j\,
\label{Zguiding}
\eeq
which is  the guiding centre $\bQ$ in the critical case \eqref{Carrcrit},
for which the equations \eqref{Cmot} imply, even for unrelated  $B^*$ and $\theta$,
\beq
\dot{\widetilde{Q}}_{i} = 0 \Rarrow \widetilde{Q}_{i} = \const
\label{noQmotion}
\eeq
Coordinate-immobility  \eqref{noxmotion} is recovered. 
when the exotic parameter is turned off, $\theta \to 0$. 

We briefly recall for comparison the discussion of the exotic Galilean model  in chiral terms \cite{ZHChiral}. The symplectic form  and thus  commutation relations are still \eqref{chirO} and \eqref{exoGchiralPB} (up to $B^* \leadsto eB$), but the Hamiltonian has a kinetic term
 which comes from  $\bp^2/2m$, 
\beq
\cH_{exoG} = \Big\{- e\bE\cdot\bX^{+} \Big\}
 \;+\; \Big\{\dfrac{e^2 B^2}{2m} (\bX^{-})^2 - e\bE\cdot\bX^{-} \Big\}\,.
\label{GchirH}
\eeq
The system splits into two uncoupled parts, however the $X_i^{-}$ equation picks up a kinetic term, cf. \eqref{Cchireq}. 
\beq\barraynb{rlr}
\displaystyle\frac{dX^{+}_{i}}{dt} &= &\;\;\epsilon_{ij}\displaystyle\frac{%
E_{j}}{B}\,,
\\[14pt]
\left(1-\theta{eB}\right)\,\displaystyle\frac{dX^{-}_{i}}{dt} &= &\; eB\epsilon_{ij}X^{-}_j -\epsilon_{ij}\displaystyle\frac{E_{j}}{B}\,.
\earraynb
\label{GX+-mot}
\eeq
For $m^*\neq0$ the guiding centre follows the Hall law and $X_i^{-}$ rotates around it.
In the critical case $m^*=0$, $X_i^{+}$ moves by following the Hall law and  $X_i^{-}$ is frozen into $X_i^{-}=E_i/eB^2$.  

For $m^*=0$ both systems become singular. For Galilei it is the frequency of the rotation wich diverges, FIG.\ref{Exotic-G}, and for Carroll it is the velocity which becomes infinite, FIG.\ref{xv-theta}.

The Souriau form of the exotic Carroll system \eqref{chirdec} splits into two parts,
\beqa
\sigma_{exoC} = 
\hskip46mm
\label{CsigmaQ}
\\[6pt]
\Big\{\half B^*\epsilon_{ij}dQ_i\wedge dQ_j +\, eE_i dQ_i\wedge ds\Big\}
&+&
\Big\{\half\big(\theta - \displaystyle\frac{1}{B^*}\big) \epsilon_{ij}dp_i\wedge dp_j +
\epsilon_{ij}\displaystyle\frac{eE_j}{B^*}\, dp_i\wedge ds\Big\}\,.\quad
\nn
\eeqa
cf. \eqref{GsigmaQ} in the Galilean case.
The first brace rules the motion of the guiding center, and the second that of the uncoupled momentum --- for any values of $B^*$ and $\theta$. 
In the singular case  $m^*=m(1-{\theta}B^*)=0$ \eqref{CsigmaQ} projects, after removing the exact $dp_i\wedge ds$ term, to 
\beq
\sigma^c_{exoC}=\half B^*\epsilon_{ij}\, dQ_i \wedge dQ_j + eE_i\, dQ_i\wedge ds
\eeq
the same as 
\eqref{xoGr} (up to $eB\leftrightarrow  B^*$ and $t \leftrightarrow s$), while the second braces are somewhat different. The trajectories are in general different: spiralling for exoG, and doubly Hall for exoC.
However in the critical case \eqref{GcritB} the Galilean form \eqref{GsigmaQ}  also decouples, and projecting into $2+1$ dimension, we get the same reduced form $\sigma_{exoC}$,
(up to  $B^* \leftrightarrow eB_{crit}$ and $s \leftrightarrow t$)
--- which is indeed  identical to the DJT form $\sigma_{DJT}$ in \eqref{DJTsigma} (up to 
$
eB \leftrightarrow B^*,\;  t \leftrightarrow s,
\;
\bQ \leftrightarrow \bx\,.
$
In conclusion~:

\bprop
Carrollian and critical Galilean motions project onto the same Hall motion, namely that for DJT in \eqref{Halllaw} (after identifying the guiding center $\bQ$ with the DJT position).
\label{C=G=DJTP}
\eprop

\section{Symmetry and immobility}\label{SymmSect}

This section is devoted to the symmetries of the various models and to 
 clarify the intimate but subtle relation between  
 boost symmetry and (im)mobility properties of Carroll particles. 
Let us keep in mind that the Hall law \eqref{Halllaw} and its anomalous exotic extension \eqref{aCHall} both imply \emph{partial immobility}: the only allowed motions are perpendicular to the electric field.

\subsection{Boost symmetry of DJT}\label{DJTboostSym}

This section is devoted to the curious \emph{double} 
 symmetry of the DJT model. We start with  Galilei boosts, 
\beq
\bx \to \bx + \bb\,t\,, \quad t \to t\,,
\label{Gboost}
\eeq
and observe that
the lhs of the Hall law \eqref{Halllaw}  is shifted,
$ 
(x_i)^{\prime} \to (x_i)^{\prime} + b_i,
$ 
whereas the rhs of \eqref{Halllaw} is left invariant. Thus~:

\goodbreak
\bprop
Galilei boost symmetry is broken for the DJT system by the electromagnetic field.
\label{DTJnoGalP}
\eprop

The statement is readily confirmed also in either the Lagrangian or in the symplectic  framework~: 
 \beqa
\cL_{DJT} &\to& \cL_{DJT} + 
\Big(-\half eB\epsilon_{ij}x_ib_jt+ \half e E_ib_i t^2  
\Big)^{\prime}+eB\epsilon_{ij}x_ib_j\,,
\\[6pt]
\sigma_{DJT} &\to& \sigma_{DJT} - eB\epsilon_{ij} b_i dx^j\wedge dt\,, 
\label{GDJGboost}
\eeqa
respectively. 

The unexpected result is that \emph{Carroll boosts},
\eqref{Cboost}, behave better~: 
 a C-boost 
  takes  the Souriau form \eqref{DJTsigma}  [up to trading Galilean time $t$ for Carroll time, $s$] into
\beq
\sigma_{DJT}  \to \sigma_{DJT} -e\epsilon_{ij}E_ib_j\,dx_1\wedge dx_2\,
\label{DJTsigmaboost}
\eeq
and the extra term  vanishes when the boost is along the electric field.
Alternatively, the DJT Lagrangian changes as, 
\beq
\cL_{DJT}ds \to \cL_{DJT} ds+ d\Big(\half eE_i x_i^2b_i\Big)
-e\Big(E_1b_2 x_1 dx_2+E_2b_1 x_2dx_1\Big)\,, 
\label{DJTLCboost}
\eeq
where the last term is a total derivative if 
$\epsilon_{ij}E_{i}b_{j}=0$. 
Thus, in contrast with the Galilean case~: 

\begin{prop}
The DJT system has a ``half-Carroll symmetry''~:
boosts which are perpendicular to the electric field are broken but those along the electric field remain unbroken.
\label{goodboostP}
\end{prop} \vskip-4mm

Conserved quantities are readily derived by the symplectic Noether theorem \cite{SSD} recalled in Appendix \ref{Appendix}. Contracting the  Souriau form \eqref{DJTsigma} by an (infinitesimal) Carroll boost $\bbeta$,
\beq
X_{Carr}: \quad \delta \bx=0 \aand \delta s = - x_i \beta_i\,,
\label{infCboost}
\eeq
 we get, choosing again $\bE = (0,E)$ for simplicity,
\beq
\sigma(X\,, \,.\,)=eE\left\{\big(x_{1}dx_{2}\big)\beta_1 + 
d\big(\frac{x_{2}^{2}}{2}\big) \beta_2\right\}\,.
\label{sigmaX}
\eeq
The 1st term here, obtained by boosting perpendicularly to $\bE$ 
i.e. in the direction of the (Hall) motion, is \emph{not} closed:
$ 
d\big(x_{1}dx_{2}\big)\beta_1  = \big(dx_{1}\wedge dx_{2}\big)\beta_1 \neq 0\, 
$. Thus~: 

\bprop
Boosting perpendicularly to $\bE$ is not a symmetry,
but boosting \underline{along $\bE$ is} a symmetry, with associated conserved quantity 
\beq
\cK_{||} = \frac{eE}{2} x_{2}^{2} 
\Rarrow x_{2} = x_{2}^0 \quad\text{fixed}\,,
\label{f2cons}
\eeq
consistently with the half-broken Carroll symmetry in
 Prop. \ref{goodboostP}. 
\label{halfDJTboostP} 
\eprop

The conservation of $\cK_{||}$ can  be checked also directly by using the equations of motion.

\smallskip
Rotational symmetry is broken when $\bE\neq0$, however the system is  invariant w.r.t. space and Carroll-time translations,
\begin{equation}
X_{C}=\gamma_{i}\frac{\partial}{\partial x_{i}}+ \varepsilon  \frac{\partial}{\partial s}\,,
\label{restCarroll}
\end{equation}%
which generate, consistently with the equations of motion, the conserved quantities
\beq
\barraynb{lcll}
p_i = &eB\epsilon_{ij}x_j + eE_i s &\qquad  &\text{Carroll momentum}
\\[4pt]
\cH = &eE_ix_i=eV(\bx) &\qquad &\text{Carroll energy}
\earraynb\quad.
\label{DJTCcons}
\eeq
 
Conversely,  momentum conservation implies the Hall law  \eqref{Halllaw} (with Carroll time), while that of the Carroll energy implies that the motion is on an equipotential. 

These  results could also by obtained in Hamiltonian terms. For example, $f_2$ Poisson-commutes with $\cH$, $\big\{f_2, \cH\big\}=0$.
Conversely, $f_2$ generates the Hamiltonian vectorfield 
$ X_1 = {E}/{B},\; X_2=0\,.$
More generally, if $f = f(x_1,x_2)$ then
$ 
\{f, \cH\} =  \frac{1}{B}\p_1f,
$ which is conserved for  an arbitrary function of $x_2$, $f = f(x_2)$.

\smallskip
It is instructive to compare the Carroll results with those in the Galilean case:
the Galilei boost symmetry is fully broken by the magnetic field, as we have seen.
Accordingly, inserting the Galilean generators into the Souriau form yields, 
\begin{eqnarray}
\sigma_{DJT} \big(X\,,\,\cdot\,\big)  
=d\Big\{\gamma_{i}e\big(B\varepsilon_{ij}x_{j}+E_{i}t\big) 
-\epsilon \big(eE_{i}x_{i}\big) + \beta_{i}\big(\frac{1}{2}eE_{i}t^{2}\big)\Big\}
+
\beta_{i} eBt\,,\epsilon_{ij}dx_{j}\,:\quad
\label{DJTnoboost}
\end{eqnarray}
the last term is not closed,
 breaking the Galilei boost symmetry, consistently with Prop.\ref{DTJnoGalP}. 
 
\subsection{Symmetries of  exotic particles}\label{exoSymmSec}

\bitem
\item 
Lifting  \eqref{Gboost}  to the evolution space of an exotic Galilean particle,
\beq
\bx \to \bx + \bb\, t,
\qquad
\bp \to \bp + {m}\bb,\qquad
t \to t,
\label{GalOnexo}
\eeq
the Souriau form \eqref{eGalsigma} changes as  for DJT in  \eqref{GDJGboost} [with $\sigma_{DJT} \leadsto \sigma_{exoG}$], leading to~:
\bprop
A nonzero magnetic field $B\neq0$
breaks Galilean boost symmetry of an exotic Galilean particle (as seen also directly from in Prop.\ref{regGalP}: the rhs is invariant w.r.t. \eqref{Gboost}).
\label{noGboostP}
\eprop 

\item
Let us now turn to exotic Carroll particles.
Assuming that  $\theta{B^*}\neq1$, the system is regular, described by the Souriau form 
 \eqref{eCarrsigma} on the evolution space $\Big\{\bx,\bp,s\Big\}$, upon which Carroll boosts act as \cite{Marsot21},%
\begin{equation}
\bx \to \bx, \qquad \bp \to \bp + {m}\bb,%
 \qquad  s \to s-b_{i}x_{i}\,,  
\label{Carrboostact}
\end{equation}%
to be compared with \eqref{GalOnexo}. 
Inserting into \eqref{eCarrsigma} we find again \eqref{DJTsigmaboost}
[up to $\sigma_{DJT}\leadsto \sigma_{exoC}$], implying: 

\bprop
For  an exotic Carroll particle,
boosts perpendicular to $\bE$ are broken by the electric field,  but  the parallel ones remain unbroken~:
we get a ``half-Carroll symmetry'' (as does its DJT reduction).
\label{halfCarrSymP}
\eprop
\eitem

\goodbreak 

Conserved quantities are readily found by  contracting the Souriau form  $\sigma_{exoC}$ with the infinitesimal boost generator \eqref{infCboost} lifted to the evolution space,
\beq
X_b: \,
\delta\bx=0\,,\quad
\delta\bp = m\bbeta\,,
\,\quad
\delta s = - \bbeta.\bx  
\label{infCboostE}
\eeq
cf. eqn. \#(3.3) of \cite{Marsot21}, yields
\beq
\sigma\,(X_b,\,\cdot\,)= d\Big\{
\beta_i(x_i +\theta \epsilon_{ij}p_j)\Big\}
+ eE_i\,dx_i\, (\beta_k x_k)\,,
\label{exoSboost}
\eeq 
which is closed when the last term is closed. But for
$\bE=(0,E)$,
\beq
\left\{\barraynb{llccll}
\text{for} & \beta_1=0\qquad
 &d\Big(\half eEx_{2}^{2} \Big)\beta_2  &\text{closed}
 &\Rarrow &\text{symmetry}
\\[8pt] 
\text{for} &\beta_2=0 &eEx_{1}dx_{2} \,\beta_1 &\text{not closed} &\Rarrow &{\text{broken}}  
 \earraynb 
\right.\qquad.
\label{sigmaXb}
\eeq
Thus, consistently with \eqref{f2cons}~: 

\bprop
Boosting the  exotic Carroll particle along the electric field $\bE=(0,E)$ generates the conserved quantity
\beq
\cK_{||}=(x_{2}-\theta\,p_1) + \half eE\,x_{2}^{2}
= \widetilde{Q}_{2} + \half eE\,x_{2}^{2}\,\,.
\label{Cbmom2}
\eeq
\label{B2P}
with the notation  in \eqref{noQmotion}.
\eprop\vskip-2mm
The  conservation of $\cK_{||}$ is readily confirmed
by using  (after replacing $eB$ by $B^*$) the commutation relations
\eqref{GexoPB}. For $\cH_{exo}=\cH=-eEx_{2}$ we find,
\beq 
\big(1-\theta{B^*}\big)\,
\Big\{\cK_{||},\cH\Big\} \propto 
\Big\{x_{2}-\theta\,p_1 + \half eE\,x_{2}^{2},eEx_{2}\Big\} \propto -{\theta} \Big\{p_1,x_{2}\Big\}=0\,.
\nn 
\eeq

Contracting $\sigma$  with $s$-translations, $\delta s = \varepsilon$, yields
$ 
\sigma(X_s,\,\cdot\;) = d\big(E_i x_i\big), 
$ 
which implies that the projection of the position onto the electric direction is a constant of the motion, 
\beq
\cE = eE_i x_i\,. 
\label{senergons}
\eeq
Thus the two terms in \eqref{Cbmom2} are separately conserved. 

At last, contracting $\sigma$ with translations,
\beq
X_{\gamma}: \quad \delta \bx = \bgamma,\quad \delta p_i = 0,\quad  \delta s = 0,
\label{sigmaXx}
\eeq
we get two conserved quantities,
\beq
\left\{\barraynb{lcccl}
\cT_1 &=&B^*\Big(x_2-\displaystyle\frac{1}{B^*}p_1\Big)
&=&B^*Q_{2}
\\[8pt]
\cT_2 &=&-B^*\Big(x_1 + \displaystyle\frac{1}{B^*}p_2
 -\displaystyle\frac{eE\,}{B^*}\,s 
 \Big)&=&-B^*\Big(Q_{1}-\displaystyle\frac{eE\,}{B^*}\,s\Big) 
\earraynb 
\right.
\label{transcons}
\eeq
which imply again that $Q_{2}$ is fixed and $Q_{1}$ follows the Hall law.
This should be compared with \eqref{GQHall} for Galilei.
\goodbreak

\begin{table}[thp]
\begin{tabular}{|c|c|c|c|c|}
  \hline   
  \text{ particle type } &\text{position} $\bx$  & guiding centre $\bQ$
  &\text {Gal-boost} & \text {Carr-boost } 
  \\
  \hline\hline  
  \text{DJT} &\text{Hall motion} 
  &--- 
  &\text{broken} &\text{half broken} 
  \\
  \hline    
  \text {Galilei} &\text{screw precession} &\text{Hall motion} &\text{broken} & ---  
  \\
  \hline    
  \text {Exotic-Galilei} &\text{anomalous screw} 
  &\text{Hall motion} & \text{broken} & ---  
  \\
  \hline 
  \text{Reduced-Galilei} & --- &\text{Hall motion} & \text{ broken} &\text{half broken} 
  \\
  \hline 
  \text{Carroll} &\text{no motion} & \text{Hall motion} 
  & --- & \text{half broken} 
  \\
  \hline 
  \text {Exotic-Carroll}  &\text {anomalous Hall motion} &\text{Hall motion} & --- & \text {half broken} \\
  \hline 
  \text {Reduced-Carroll} & --- &\text{Hall motion} & \text{broken} &\text{half broken} \\
  \hline
\end{tabular}
  \caption{\textit{\small Motions and boosts symmetries.} }
\end{table}

DJT is the commun projection of
both the exo-Galilei and exo-Carroll systems  (up to $t \leftrightarrow s$) 
which thus and inherits both of their symmetries:
 Galilei boost symmetry is broken, however Carroll boosts along the electric field are symmetries.

\goodbreak

\section{Motion on the Black Hole horizon}\label{BHhor}

Our results above are analogous to but different from those we had found before on the horizon of a black hole for an exotic Carroll particle with parameters $\kappa_{exo}$ and $\kappa_{mag}$, with
 \emph{zero electric charge} but with a magnetic moment $\mu$ and anyonic spin $\chi$ \cite{Marsot21,MZH,Gray}. Its Hamiltonian is $\cH=-(\mu\chi)B$, where $B$ may {not} be a constant (as it happens for a Kerr-Newman black hole). The equations of motion are, 
 \beq
\dot{x}_i = (\mu\chi)\frac{\kappa_{exo}}{\kappa_{exo}\kappa_{mag}-m^{2}}
\epsilon_{ij}\p_jB\,,
\label{MZHeq}
\eeq
cf. eqn. \#(2.4) of \cite{MZH}.
 The product $(\mu\chi)$ behaves as an electric charge and  the magnetic field, $B$,  plays a role analogous to that of an electric potential, cf. \eqref{CHamilt}. 
  
For non-zero mass $m$, putting  again $\theta = \kappa_{exo}/m^2$ we get an \emph{anomalous Hall effect} for the position, 
\beq
\dot{x}_i = -(\mu\chi)\frac{\theta}{1-\theta\kappa_{mag}}
\epsilon_{ij}\p_jB\,,
\label{exophoton}
\eeq
which is a sort of dual to \eqref{anomHallef}
 with the gradient of the magnetic field behaving as an electric field. 
 Turning off the exotic structure, $\theta \to 0$, the particle \emph{stops moving} and we recover the celebrated \emph{immobility of massive unextended Carroll particles}. 

Letting instead the mass go to zero (as suggested by DJT in the Peierls case) while keeping $\kappa_{exo}\neq0$, the latter drops out, leaving us with,
\beq
\dot{x}_i = (\mu\chi)\epsilon_{ij}\frac{\p_jB}{\kappa_{mag}}\,.
\label{exophotonm0}
\eeq
Our ``exotic Carroll photons'' thus exhibits an \emph{anomalous spin-Hall effect} on the black hole horizon, underlining the ``duality" 
of $(\mu\chi)$  with electric charge, the gradient of the magnetic field playing the role of  an electric field, and second Carroll charge $\kappa_{mag}$ behaving  as an effective magnetic field,
\beq
e \leadsto  \mu\chi\,,
\qquad
\bE \leadsto \bnabla B\,,
\qquad
B \leadsto \kappa_{mag}\,.
\label{dualexo}
\eeq
 Magnetic fields which differ by a constant generate identical motion. A constant magnetic field $B=\const$ would, in particular, imply immobility, --- and it precisely our search for non-trivial motion that  led us to Kerr-Newman BHs \cite{MZH}.

\section{Conclusion}
 
The Peierls-Dunne-Jackiw-Trugenberger system \eqref{DJTlag} \cite{DunneJP,DJT}, which resurrects Peierls' historical idea \cite{Peierls}, can be derived \emph{both} from the doubly extended (``exotic'') Galilean \cite{DHPeierls,LLGal,DHJPA,NCLandau,ExoRev,ZHChiral}, \emph{and} from the recently proposed exotic Carroll model \cite{Azcarraga,Ancille0,Ancille1,Ancille2,Marsot21,MZH}
by Hamiltonian reduction \cite{FaJa}.
 
The Galilean case is well-known, therefore we focus our attention at the Carroll case. 
Having suppressed the kinetic term from the Hamiltonian results in a substantially poorer dynamics: the equations for position and for momentum are fully decoupled and the position performs \emph{anomalous Hall motion} which, for particular values of the parameters, becomes ordinary Hall motion. The extended model allows for an additional, constant ``internal'' magnetic field represented by our parameter $\kappa_{mag}$ which is added to the external magnetic field $B$ \eqref{B*field} which might play a role in the anomalous Hall effect \cite{Kar+} with no external field.

Neither the anomalous  and let alone the usual Hall effect can be obtained by an ``ordinary'' [meaning unextended] Carroll particle.
By \eqref{aCHall}  $\dot{\bx}\neq 0$ requires  $\theta \neq0$. The  condition  
\eqref{CarrHall} requires the constraint \eqref{thetakappa1} which in the massive case $m\neq0$  can not be satisfied when either of the ``exotic'' parameters vanishes.
When the exotic parameter is turned off, 
$\theta \to 0$, then eqn \eqref{aCHall} behaves as
$ 
\dot{x}_i \to 0
$ 
confirming the general wisdom that (unextended)  \emph{Carroll particles do not move}.

Recent interest in Carroll symmetry comes also from the study of \emph{fractons} in condensed matter physics \cite{Pretkofractons,PretkoCY,Gromov,Seiberg, DoshiG, Bidussi,JainJensen,Grosvenor:2021hkn,Surowka,MZCH}, whose  
 restricted mobility was attributed to their \emph{conserved dipole momentum} ---  generated by a ``dual'' of Carroll symmetry \eqref{Cboost}  familiar from the BMS context \cite{Bagchi,BMSCarr,Bagchi22,LauraPR}.

Putting a DJT particle into a planar electromagnatic field  breaks Galilei boost symmetry (Proposition \ref{DTJnoGalP}). However, owing to the poorer nature of Carroll dynamics, one of the Carroll boosts remains a symmetry for  DJT  (Proposition \ref{goodboostP})~: we get what we called  half-Carroll symmetry.

The DJT dynamics actually follows from that of exotic  Carroll particles, see Proposition  \ref{halfCarrSymP}~:
half-boost symmetry implies the conservation law \eqref{Cbmom2}   for $\cK_2$, (Proposition \ref{B2P}) which, combined with the immobility of $\widetilde{Q}$ in \eqref{noQmotion}, implies in turn no motion along 
$\bE$. 
$\bE=0$ has arbitrary direction yielding no motion at all.

The ``no-motion-property'' has recently reemerged, independently, in condensed matter physics in the study of quasiparticles called fractons \cite{Pretkofractons,PretkoCY,Gromov,Seiberg, DoshiG, Bidussi,JainJensen,Grosvenor:2021hkn,Surowka} whose restricted mobility is attributed to their \emph{conserved dipole momentum} -- generated   by a dual counterpart of Carroll symmetry \cite{MZCH}. 

Particles with no electric charge but with magnetic momentum and anyonic spin which move on the horizon of a black hole \cite{MZH,Gray} behave also analogously to our electrically charged Carroll particles. 

Our study shows strong similarity also with vortex dynamics \cite{Coslab} as it will be explained in a forthcoming paper \cite{ZenVort}.
In conclusion, 450 years after Galilei \cite{GalileiDialogo,Gassendi,Iglesias}
the Carrollian context rekindles once again the subtle question of boosts.
 
\begin{acknowledgments}\vskip-4mm
We are indebted to X. Bekaert, J.~Balog, M. Elbistan, L.~Feh\'er, G.~Gibbons, P.~Iglesias and F.~Ziegler for discussions and for correspondence.
PMZ was partially supported by the National Natural Science Foundation of China (Grant No. 11975320). 
\end{acknowledgments}
\goodbreak



\appendix
\section{\bf The Noether theorem in Souriau's framework
}\label{Appendix}

Here we summarise Souriau's version of the Noether theorem, taken from chapter 12 of \cite{SSD}. First 
we recall that a mechanical system is described by a closed 2-form $\sigma$ of constant rank defined on an evolution space $\cE$. The motions $\gamma$ are the characteristic curves of $\sigma$,
\beq
\sigma({\gamma}^{\prime}\,, \cdot \;) = 0 \,.
\label{Smotion}
\eeq
A transformation $\Phi$ of the evolutions space is a symmetry if it leaves $\sigma$ invariant,
\beq
\Phi^*\sigma = \sigma\,.
\label{sigmasymm}
\eeq
Infinitesimally, 
a vector field $X$ is a symmetry when it preserves $\sigma$; which boils down to that 
contracted by $X$ is the differential of a function $f_X$,
\beq
\sigma(X,\cdot\,)= df_X \,; 
\label{SsymmCond}
\eeq
then $f_X$ is a constant of the motion. 
The Souriau version is  equivalent to the more conventional approaches. The Lagrangian should, for example, change by a total derivative. 
The relation with the Hamiltonian framework  is  obtained in turn by splitting the Souriau form as, 
\beq
\sigma = \Omega - d\cH \wedge dt\,.
\label{sigmaOmegaH}
\eeq
The Poisson Bracket (PB) associated with the symplectic form $\Omega$  
is :
\beq
\Big\{f, g\Big\} = \Omega^{ij}\Big(\p_{i}f \,\p_{j}g-\p_{j}f\,\p_{i}g\Big)\,,
\label{OmegaPB}
\eeq
where $\Omega^{ij}$ is the inverse of the symplectic matrix $\Omega_{ij},\,\Omega^{ij}\Omega_{jk}= \delta^{i}_{\,k} $. For $f$ may depend explicitly on $t$, then the symmetry condition \eqref{SsymmCond}
requires that 
\beq
\p_tf +\Big\{f,\cH\Big\}=0\,.
\label{PBcons}
\eeq

\end{document}